\documentclass[aps,prb,twocolumn,floatfix,superscriptaddress]{revtex4}
\usepackage{graphicx}
\usepackage{amsmath}
\usepackage{hyperref}
\usepackage{bm}
\usepackage[dvips]{color}

\graphicspath{{figs/},
      }

\newcommand{\degr}{$^{\circ}$}
\newcommand{\etal} {\textit{et al.}}
\newcommand{\ie} {\textit{i.e.}}

\newcommand{\eg} {\textit{e.g.}}

\newcommand{\fzd} {Institute of Ion Beam Physics and Materials Research, Forschungszentrum Dresden-Rossendorf, P.O. Box 510119, 01314 Dresden, Germany}

\newcommand{\seu} {Department of Physics, Southeast University, Nanjing 211189, China}
\newcommand{\als} {Advanced Light Source, Lawrence Berkeley National Laboratory,
Berkeley, California 94720, USA}

\newcommand{\NiFeO} {NiFe$_2$O$_4$}
\newcommand{\CoFeO} {CoFe$_2$O$_4$}

\begin{document}
\title{Spinel ferrite nanocrystals embedded inside ZnO: magnetic, electronic and magneto-transport properties}

\date{\today}

\author{Shengqiang~Zhou}
\email[Electronic address: ]{S.Zhou@fzd.de} \affiliation{\fzd}
\author{K.~Potzger}
\affiliation{\fzd}
\author{Qingyu Xu}
\affiliation{\seu}
\author{K.~Kuepper}
\affiliation{\fzd}
\author{G.~Talut}
\author{D.~Mark\'{o}}
\affiliation{\fzd}
\author{A.~M\"{u}cklich}
\author{M.~Helm}
\author{J.~Fassbender}
\affiliation{\fzd}
\author{E.~Arenholz}
\affiliation{\als}
\author{H.~Schmidt}
\affiliation{\fzd}

\begin{abstract}
In this paper we show that spinel ferrite nanocrystals (\NiFeO,
and \CoFeO) can be texturally embedded inside a ZnO matrix by ion
implantation and post-annealing. The two kinds of ferrites show
different magnetic properties, \eg~coercivity and magnetization.
Anomalous Hall effect and positive magnetoresistance have been
observed. Our study suggests a ferrimagnet/semiconductor hybrid
system for potential applications in magneto-electronics. This
hybrid system can be tuned by selecting different transition metal
ions (from Mn to Zn) to obtain various magnetic and electronic
properties.
\end{abstract}
\maketitle

\section{Introduction}\label{introduction}

Spinel ferrites are materials with rich magnetic and electronic
properties \cite{shikazumi}. As bulk materials, they can be
half-metallic (such as Fe$_3$O$_4$) or insulating (most spinel
ferrites), ferrimagnetic (most spinel ferrites) or
anti-ferromagnetic (ZnFe$_2$O$_4$). Insulating ferrites (such as
\NiFeO~and ZnNiFe$_2$O$_4$) are usually referred to as magnetic
insulators. These kinds of materials are technologically important
with various applications as permanent magnets, microwave devices,
and magnetic recording media. Physically the magnetic and
electronic properties of spinel ferrites are determined by the
cation distribution among the tetrahedral (A) and octahedral (B)
sites. The growth of low-dimensional spinel ferrites of both thin
films and nanoparticles has shown the possibility to tune the
cation distribution, therefore resulting in magnetic and
electrical properties drastically different from bulk materials.
L\"{u}ders \etal~have shown that the conductivity of NiFe$_2$O$_4$
thin films can be tuned over five orders of magnitude by varying
the growth atmosphere \cite{luders_AM}. The sites of Fe$^{3+}$ can
be changed from A to B sites in ZnFe$_2$O$_4$ nanoparticles,
resulting in ferrimagnetism \cite{zhou07JPD}. Geiler
\etal~proposed a method to design and control the cation
distribution in hexagonal BaFe$_{12-x}$Mn$_x$O$_{19}$ ferrites at
an atomic scale, which results in the increase of magnetic moment
and N\'{e}el temperature \cite{geiler:067201}. Moreover, most of
transition metals (TM) can form solid solutions with Fe$_3$O$_4$,
resulting in TM$_x$Fe$_{3-x}$O$_4$ spinel alloys with x ranging
from 0 to 1, which provides an additional degree of freedom to
tune their magnetic and electronic properties
\cite{ishikawa_APL,takaobushi_APL}. In previous research,
anomalous Hall effect and magnetoresistance have been found in
spinel ferrite thin films or granules at room temperature,
demonstrating spin-polarization of free carriers. Moreover,
ferrite thin films \NiFeO~and \CoFeO~with different conductivities
have been demonstrated to be useful as electrodes or spin-filter
in magnetic tunnel junctions
\cite{luders_AM,Luders_NiFeO_APL,ramos:122107,snoeck:104434,ramos:180402}.
However, to our knowledge, very limited effort has been spent to
integrate ferrite oxides with semiconductors. The growth of
ferrite oxides requires high temperatures and oxygen environment,
which is detrimental to conventional semiconductors like Si and
GaAs \cite{chen_APL_ferrites}. This explains why oxide insulators
such as MgO, Al$_2$O$_3$, and SrTiO$_3$ are mostly used as the
substrates to grow ferrite oxides \cite{suzuki01}. In this paper,
we show that TMFe$_2$O$_4$ nanocrystals (TM=Ni, Co) can be
embedded inside ZnO, and we present a systematic study on their
magnetic, electronic, and transport properties. The various
ferrites with different magnetic properties synthesized inside a
semiconducting matrix open a new avenue for fabricating hybrid
systems.



\section{Experiments}\label{experiments}

We utilize different methods to characterize the ferrite/ZnO
hybrid systems. The aim is to show the similarity in structure,
but variability in magnetic, electronic and magneto-transport
properties.

Commercial ZnO(0001) single crystals with the thickness of 0.5 mm
from Crystec were co-implanted with $^{57}$Fe and Ni or Co ions at
623 K with a fluence of $4\times10^{16}$ and $2\times10^{16}$
cm$^{-2}$, respectively. The implantation energy was 80 keV for
all three kinds of elements. This energy resulted in the projected
range of $R_P=38, 37, 37$ nm, and the longitudinal straggling of
${\Delta}R_P=17, 17, 17$ nm, respectively, for Fe, Co, and Ni.
Therefore, the implanted Fe ions are in the same range as the Co
and Ni ions. The maximum atomic concentration is about 10\% and
5\% for Fe and Ni(Co), respectively (TRIM code \cite{trim}). The
maximum implanted depth is around 80 nm (around 5\% of the maximum
concentration) from the surface. Annealing was performed in a high
vacuum (base pressure $\leq$10$^{-6}$~mbar) furnace at 1073 K for
60 minutes. In our previous study we have performed detailed
annealing investigation for transition metal implanted ZnO single
crystals \cite{zhou07JPD,zhou07CoNi,zhouFe}. Briefly, more
metallic clusters formed when annealing at mild temperatures (823
K or 923 K), while the oxidation starts at around 1073 K. Keeping
this high temperature, a longer annealing time results in the
formation of ferrites in Fe implanted ZnO.

Magnetic properties were measured with superconducting quantum
interference device (SQUID, Quantum Design MPMS) magnetometery.
The samples were measured with the field along the sample surface.
The temperature dependent magnetization measurement was carried
out in the following way. The sample was cooled in zero field from
above room temperature to 5 K. Then a 50 Oe field was applied, and
the zero-field cooled (ZFC) magnetization curve was measured with
increasing temperature from 5 to 350 K, after which the
field-cooled (FC) magnetization curve was measured in the same
field from 350 to 5 K with decreasing temperature.

Structural analysis was performed by synchrotron radiation x-ray
diffraction (SR-XRD) and transmission electron microscopy (TEM,
FEI Titan). SR-XRD was performed at the Rossendorf beamline (BM20)
at the ESRF with an x-ray wavelength of 0.154 nm. The
cross-section specimen for TEM investigation was prepared by the
conventional method including cutting, glueing, mechanical
polishing, and dimpling procedures followed by Ar$^+$ ion-beam
milling until perforation. The ion-milling was performed using a
"Gatan PIPS". The milling parameters were: 4 keV, 10 $\mu$A ion
current at milling angle of 4\degr~with respect to the specimen
surface. The area around the hole is electron-transparent
(thickness$<$100 nm).

Element-specific electronic properties were investigated by X-ray
absorption spectroscopy (XAS) and X-ray magnetic circular
dichroism (XMCD) at the Fe, Co and Ni L$_{2,3}$ absorption edges.
These experiments were performed at beamlines 8.0.1 (XAS) and
6.3.1 (XMCD) of the Advanced Light Source (ALS) in Berkeley,
respectively. Both total electron yield (TEY) and total
fluorescence yield (TFY) were recorded during the measurement.
While TFY is bulk sensitive, TEY probes the near-surface region.
For XMCD, the measurement was done at the minimum achievable
measurement temperature of 23 K in TEY mode. A magnetic field of
$\pm$2000 Oe was applied parallel to the beam. The grazing angle
of the incident light was fixed at 30\degr~with respect to the
sample surface.

Conversion electron M\"ossbauer spectroscopy (CEMS) in
constant-acceleration mode at room temperature (RT) was used to
investigate the Fe lattice sites, electronic configuration and
corresponding magnetic hyperfine fields. The spectra were
evaluated with Lorentzian lines using a least squares fit
\cite{brand87}. All isomer shifts are given with respect to
$\alpha$-Fe at RT.

Magnetotransport properties were measured using Van der Pauw
geometry with a magnetic field applied perpendicular to the film
plane. Fields up to 60 kOe were applied over a wide temperature
range from 5 K to 290 K and the carrier concentration and the
majority carrier mobility were extracted. 

\section{Results and discussion} \label{results}

\subsection{Structural properties}

\begin{figure} \center
\includegraphics[scale=0.8]{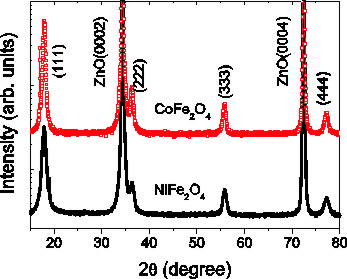}
\caption{SR-XRD 2$\theta$-$\theta$ scan revealing the formation of
\NiFeO~and \CoFeO~in (Ni, Fe) or (Co, Fe) co-implanted ZnO. In
both pattern, the diffraction peaks of (111), (222), (333) and
(444) from ferrites are clearly visible. The diffraction peaks
(0002) and (0004) from ZnO are also indicated. The small and sharp
peaks at the left side of \CoFeO(111) and the right side of
\NiFeO(111) cannot be identified at this stage. The peak at the
left side of \CoFeO(111) could be the forbidden peak of ZnO(0001),
which appears due to the lattice damage. However, another
forbidden peak of ZnO(0003) does not show up. Note that the two
peaks in the two spectra are not at the same angular position, and
both correspond to very large lattice distances. Some noise-like
peaks are also shown in other paper \cite{bonanni:135502} and
could not be identified.}\label{fig:XRD_NiCoFeO}
\end{figure}

\subsubsection{X-ray diffraction}

Figure \ref{fig:XRD_NiCoFeO} shows the SR-XRD patterns for the
annealed samples. Besides the strong peaks from ZnO(0002) and
(0004), four small peaks arise for each sample. They are assigned
to (111), (222), (333) and (444) diffractions for \NiFeO~and
\CoFeO, respectively. This implies that these nanocrystals are
(111) textured inside the ZnO
matrix. 
However, some nanocrystals with (400) orientation have been also
observed by TEM as shown below.
The crystallite size is estimated using the Scherrer formula
\cite{scherrer}.
\begin{equation}\label{scherrer}
    d=0.9\lambda/(\beta\cdot\cos\theta),
\end{equation} where $\lambda$ is the wavelength of the X-ray, $\theta$ the Bragg angle, and $\beta$ the FWHM of
2$\theta$ in radians. The average crystallite size is deduced to
be around 12 nm and 15 nm for \NiFeO~and \CoFeO~nanocrystals,
respectively.

\subsubsection{TEM}

\begin{figure} \center
\includegraphics[scale=0.3]{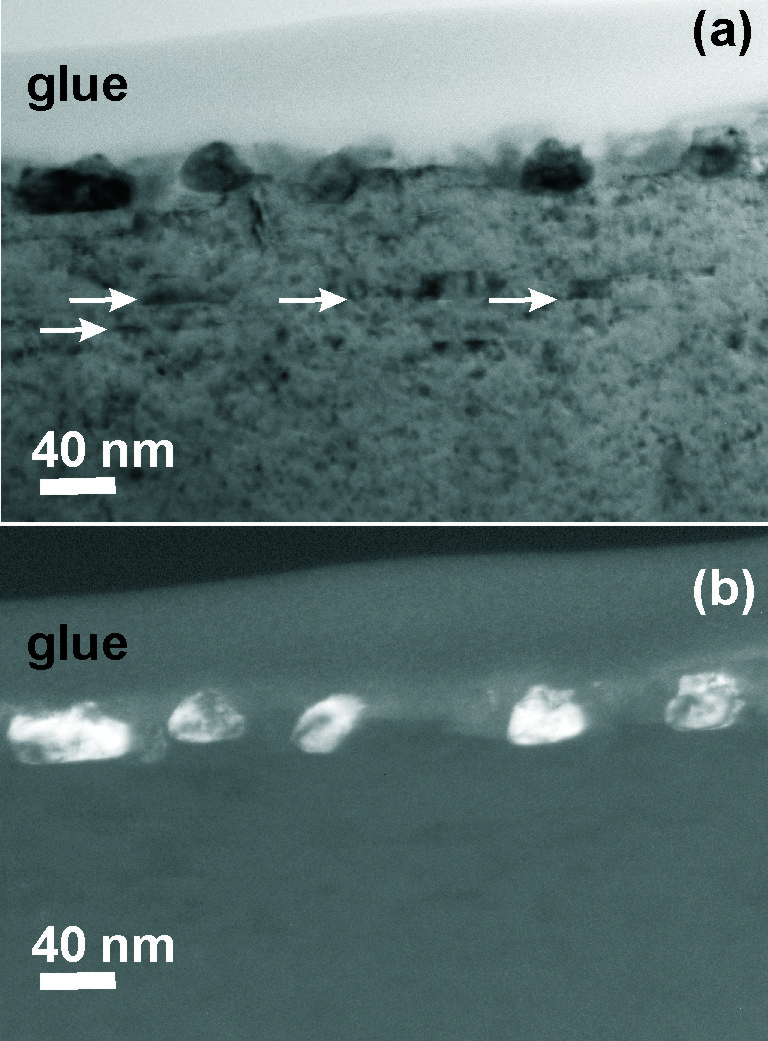}
\caption{Cross-section TEM image of Fe and Ni co-implanted ZnO
after annealing (a) bright field and (b) dark
field.}\label{fig:TEM_lm}
\end{figure}

In order to confirm the formation of ferrite nanocrystals, high
resolution cross-section TEM was performed on selected samples.
Fig. \ref{fig:TEM_lm}(a) displays the bright-field TEM images. In
an overview, there are three features. The grains of secondary
phases are located in the surface region, which are identified as
\NiFeO. Some planar extended defects are indicated by arrows, and
are parallel to the basal plane of the ZnO wurtzite structure in a
depth of around 60 nm. These extended defects are caused by ion
implantation in ZnO \cite{kucheyev03} and are usually populated at
the end of the ion range. The third feature is the dark-spot,
which is also located in the depth of unimplanted ZnO. The
formation of \NiFeO~at the near-surface depth is also confirmed by
the dark-field TEM image as shown in Fig. \ref{fig:TEM_lm}(b) of
the same area of Fig. \ref{fig:TEM_lm}(a). The out-diffusion of Fe
upon annealing at high temperatures has been observed in ZnO
\cite{Zhou2009} as well as in TiO$_2$ \cite{zhou07tio2}.

\begin{figure} \center
\includegraphics[scale=0.41]{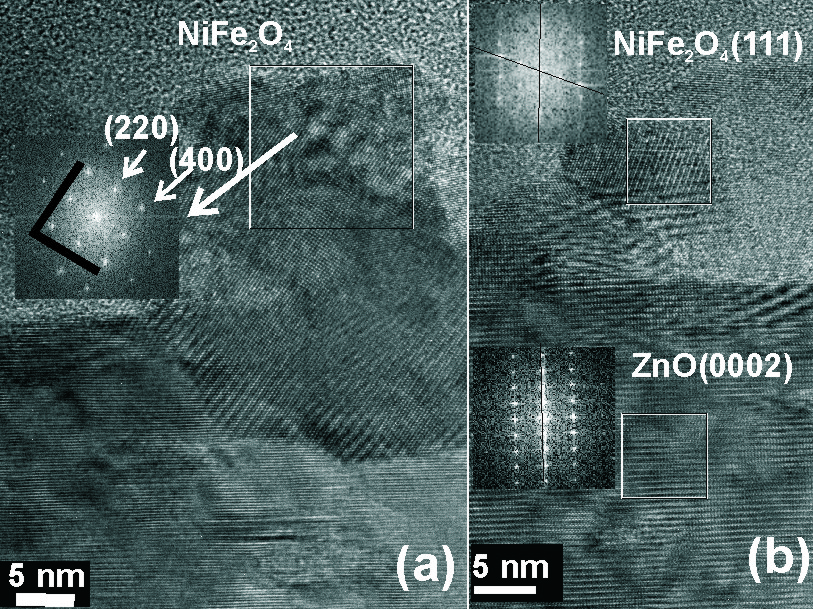}
\caption{High resolution TEM image for representative
\NiFeO~nanocrystals. (a) the specimens is tilted by 11\degr.
\NiFeO~nanocrystal is identified. The black lines guide the eyes
to show the cubic symmetry of the secondary phase. (b) Another
\NiFeO~nanocrystals with the orientation of
(111)$\parallel$ZnO(0001) as confirmed by FFT patterns. The
clearly visible planes are \NiFeO($\overline{1}$11) with an angle
of $\sim$ 71\degr~from the surface. In FFT patterns the dashed
lines indicate the sets of lattice planes. }\label{fig:TEM_NiFeO}
\end{figure}

Using high resolution TEM we identified the secondary phase to
confirm the XRD results. As shown in Fig. \ref{fig:TEM_NiFeO}(a),
the sample was tilted in order to have a better view on the
nanocrystals. Note that the lattice planes are more clear in the
nanocrystals than that in the ZnO substrate. The inset of Fig.
\ref{fig:TEM_NiFeO}(a) is the Fast Fourier Transform (FFT) of the
image indicated by a square. The FFT clearly shows the
cubic-symmetry of the nanocrystal. The two sets of lattice
spacings amount to 0.291 nm and 0.207 nm, and correspond to
\NiFeO(220) and (004), respectively. Concerning the orientation
between \NiFeO~and the ZnO matrix, XRD in general provides the
integral information over a large area of the sample, while TEM is
a rather localized method. By high resolution TEM, we found some
grains with [111] orientation as shown in Fig.
\ref{fig:TEM_NiFeO}(b). By FFT two sets of lattice planes are
identified to \NiFeO$\{111\}$. One is parallel with the sample
surface, while the other is around 71\degr~away from the surface
[ZnO(0001)]. This is in agreement with a fcc structure of \NiFeO.
However there are also some grains with [001] orientation, \eg~the
one in Fig. \ref{fig:TEM_NiFeO}(a). One also can see some moire
fringes in the ZnO part due to the overlap of \NiFeO~and ZnO.

Note that the \NiFeO~grains (see Fig. \ref{fig:TEM_lm}) are as
large as 20-40 nm, and larger than the values determined from XRD.
However, one grain does not have to correspond to one
\NiFeO~nanocrystal. On the other hand, in the dark-field image all
the grains show similar sizes as that in the bright field. This is
due to the fact that these nanocrystals are well oriented. By high
resolution TEM we examined more than 10 nanocrystals in different
areas of the specimens. Their diameters are in the range of 10-20
nm, which is in a qualitative agreement with the XRD measurement.

\subsection{Magnetic properties}\label{sec:mag}

\begin{figure} \center
\includegraphics[scale=0.62]{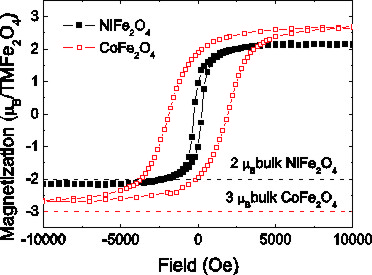}
\caption{Hysteresis loops measured at 5 K for \NiFeO~and
\CoFeO~nanocrystals. They show drastic difference in coercivity
field.}\label{fig:MH_NiCoFeO}
\end{figure}

By structural analysis, we have shown the formation of \NiFeO~and \CoFeO~nanocrystals inside the ZnO matrix. In
this section, we will compare their magnetic properties. Fig. \ref{fig:MH_NiCoFeO} shows the hysteresis loops
measured at 5 K. The differences between \NiFeO~and \CoFeO~are significant. At 5 K, the coercivity of \CoFeO~is
1900 Oe, and much larger than the coercivity of \NiFeO~amounting to 280 Oe, \ie~one is a hard magnet, and the
other is a soft one. For comparison the saturation magnetization of bulk crystals is also indicated in Fig.
\ref{fig:MH_NiCoFeO}. \NiFeO~nanocrystals have a slightly larger value than bulk \NiFeO, and a smaller value for
\CoFeO~nanocrystals. This could be due to the cation site exchange between Ni$^{2+}$ (Co$^{2+}$) and Fe$^{3+}$
and will be discussed in section \ref{electronic}. However, another possibility is that there is a mixture of
Ni$^{2+}$ and Fe$^{2+}$ at tetrahedral sites resulting in (Ni,Fe)Fe$_2$O$_4$ (Ni$_{1-x}$Fe$_{2+x}$O$_4$). The
magnetization for bulk Fe$_3$O$_4$ is 4.1 $\mu_B$ per formula unit. To verify this, one needs to perform a
precise local Fe, Ni(Co) concentration measurement. We performed an electron energy-loss spectroscopy (EELS)
analysis to profile the composition of the ferrite nanocrystals during the TEM measurements. We could not probe
an elementally resolvable EELS signal possibly due to the similar atomic number of the embedding ZnO matrix and
the ferrite, given the complex element types (Fe, Co/Ni and Zn) within the probe area. Macroscopically, the
appearance of Fe, Co, Ni is clearly revealed by x-ray absorption as shown later in Sec. \ref{electronic}. In
literature the application of EELS in similar cases (embedded nanocrystals) is mainly for qualitative
investigation \cite{li:131919,Wang20092291,opel}. On the other hand, exposing the nanocrystals to the electron
beam for a longer time results in a heavy beam damage and contamination of the specimens, as well as the
structural modification of the nanocrystals \cite{rellinghaus}.

\begin{figure} \center
\includegraphics[scale=0.62]{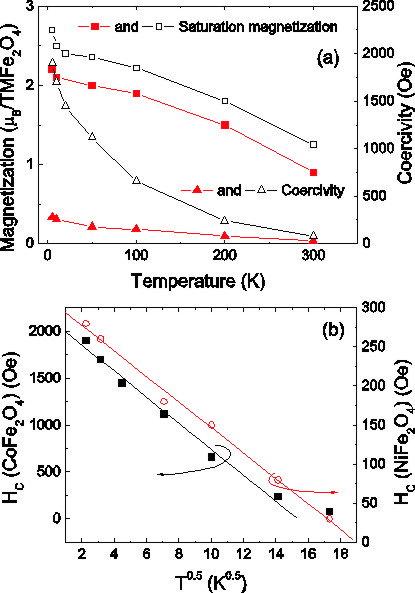}
\caption{(a) Temperature dependent saturation magnetization and
coercivity for \NiFeO~(red solid symbols) and \CoFeO~(black open
symbols). The solid lines are guides for eyes. (b) The plot of
coercivity as a function of T$^{1/2}$. }\label{fig:T_MT_NiCoFeO}
\end{figure}

Fig. \ref{fig:T_MT_NiCoFeO} shows the temperature dependent
saturation magnetization and coercivity. One sees clearly that the
coercivity decreases exponentially with increasing temperature.
This is expected for a magnetic nanoparticle system. According to
the Stoner-Wohlfarth theory \cite{stoner}, the magnetic anisotropy
energy E$_A$ of a single domain particle can be expressed as:
\begin{equation}\label{stoner} E_A = KVsin^{2}\theta,
\end{equation} where K is the magnetocrystalline anisotropy
constant, V the volume of the nanoparticle, and $\theta$ is the
angle between the magnetization direction and the easy axis of the
nanoparticle. This anisotropy serves as the energy barrier to
prevent the change of magnetization direction. When the size of
magnetic nanoparticles is reduced to a critical value, E$_A$ is
comparable with thermal activation energy, $k_BT$, the
magnetization direction of the nanoparticle can be easily moved
away from the easy axis by thermal activation and/or an external
magnetic field. The coercivity of the nanoparticles is closely
related to the magnetic anisotropy. At a temperature below
blocking temperature T$_B$, the coercivity corresponds to a
magnetic field which provides the required energy in addition to
the thermal activation energy to overcome the magnetic anisotropy.
As temperature increases, the required magnetic field (H$_C$) for
overcoming the anisotropy decreases. At the temperature of 0 K,
where all the magnetic moments are blocked, the coercivity is
equal to the value for single domains. At a high enough
temperature, when all moments fluctuate with a relaxation time
shorter than the measuring time, coercivity equals zero. In the
temperatures between the two extremes the coercivity H$_C$ can be
evaluated by the following formula \cite{candela:868}:
\begin{equation}\label{stonerHc}
    H_C = H_{C0}[1-(\frac{T}{T_B})^{1/2}],
\end{equation} where $H_{C0}$ is the coercivity at 0 K, and T$_B$
the blocking temperature. Fig. \ref{fig:T_MT_NiCoFeO}(b) shows a
plot of H$_C$ as a function of T$^{1/2}$. For the \NiFeO~system
$H_C$ roughly obeys a linear dependence on $T^{1/2}$ in the whole
measured temperature range. The deduced blocking temperature lies
around 360 K, which is rather close to the value found by the
ZFC/FC magnetization as shown below. The poor fitting for the
\CoFeO~system may be due to the fact that the magnetocrystalline
anisotropy energy of \CoFeO~is much larger (two orders of
magnitude) than \NiFeO. For a similar grain size the blocking
temperature of \CoFeO~can be much higher for \NiFeO. In such a
case, the measured temperature range is not large enough compared
to the high blocking temperature. This results in a large error in
the fitting.

\begin{figure} \center
\includegraphics[scale=0.8]{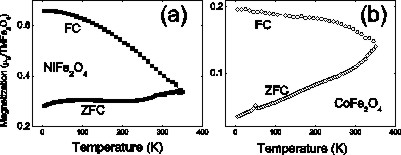}
\caption{ZFC/FC magnetization curves measured with a field of 50
Oe. (a) \NiFeO~and (b) \CoFeO. Up to 350 K, no ZFC maximum was
observed for \CoFeO.}\label{fig:ZFCFC_NiCoFeO}
\end{figure}

Fig. \ref{fig:ZFCFC_NiCoFeO} shows the ZFC/FC magnetization curves
measured at 50 Oe. An irreversible behavior is observed in ZFC/FC
curves. Such an irreversibility originates from the anisotropy
barrier blocking of the magnetization orientation in the
nanoparticles cooled under zero field. The magnetization direction
of the nanoparticles is frozen as the initial status at high
temperature, \ie, randomly oriented. At low temperature (5 K in
our case), a small magnetic field of 50 Oe is applied. Some small
nanoparticles with small magnetic anisotropy energy flip along the
field direction, while the large ones do not. With increasing
temperature, the thermal activation energy together with the field
flips the larger particles. This process results in the increase
of the magnetization in the ZFC curve with temperature. The size
distribution of nanoparticles, \ie~the magnetic anisotropy is
usually not uniform in the randomly arranged nanoparticle systems.
The larger the particles, the higher the E$_A$, and a larger
$k_BT$ is required to become superparamagnetic. The gradual
increase and the small upturn at around 20 K in the ZFC curves is
due to the size distribution of nanocrystals. In the ZFC curve for
\NiFeO~[Fig. \ref{fig:ZFCFC_NiCoFeO}(a)] a broad maximum is
observed at around 330 K, while for \CoFeO~[Fig.
\ref{fig:ZFCFC_NiCoFeO}(a)] no maximum can be seen up to 350 K.
The mean blocking temperature for \CoFeO~is well above room
temperature, which is evidenced also from the rather large
coercivity field of 80 Oe at 300 K (see Fig.
\ref{fig:T_MT_NiCoFeO}). Note that the ZFC/FC magnetization of
\NiFeO~is much smaller than that of \CoFeO. This is due to the
much larger coercivity field (magnetocrystalline constant K) of
\CoFeO, which is well above the small applied field of 50 Oe.

Since the blocking temperature is closely related to the magnetic
anisotropy energy E$_A$, one can evaluate the size of nanomagnets
by the measured T$_B$. For a dc magnetization measurement in a
small magnetic field by SQUID magnetometry, T$_{B}$ is given by
\begin{equation}\label{blocking_squid}
    T_{B,Squid}\approx\frac{KV}{30k_B},
\end{equation} where $K$ is the anisotropy energy density,
$V$ the particle volume, $k_B$ the Boltzmann constant
\cite{respaud}. $K$ is 6.3$\times$10$^3$ and 4.0$\times$10$^5$
Jm$^{-3}$ for bulk \NiFeO~and \CoFeO, respectively, at room
temperature \cite{PhysRev.80.744,PhysRev.99.1788}. Due to its
large magnetocrystalline anisotropy the maximum in ZFC curve of
\CoFeO~nanocrystals cannot be seen within the measured temperature
range. That means the blocking temperature is much larger than 350
K, which corresponds to an average diameter of \CoFeO~larger than
9 nm if assuming the value of $K$ for a bulk \CoFeO. For
\NiFeO~nanocrystals, $K$ is much smaller. Therefore, we can see a
maximum at around 320 K in the ZFC magnetization curve. Using the
$K$ value for bulk \NiFeO, the average diameter of \NiFeO~can be
calculated and amounts to 34 nm. This value, however, is larger
than that deduced from XRD and TEM measurements. The large
discrepancy is resulting from the underestimation of $K$ by
assuming the value of a bulk crystal. $K$ can be largely enhanced
due to strain, and surface effect in \NiFeO~nanomagnets, but
relatively less enhanced in \CoFeO~\cite{guogy3M}. The later has
been confirmed in strained epitaxial \CoFeO~thin films
\cite{suzuki:714}.

\subsection{Electronic configuration}\label{electronic}

\subsubsection{X-ray absorption spectra}

\begin{figure} \center
\includegraphics[scale=0.8]{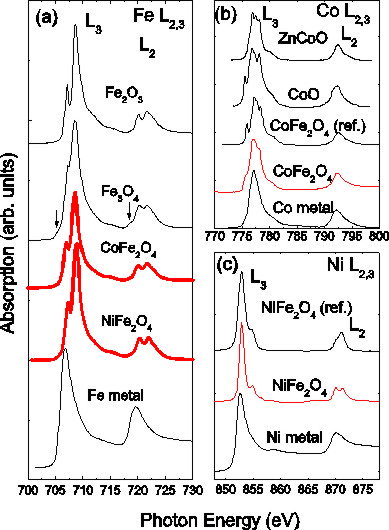}
\caption{XAS of \NiFeO and \CoFeO~along with reference spectra
from pure metal and oxides at the (a) Fe L$_{2,3}$-edge, (b) Co
L$_{2,3}$-edge and (c) Ni L$_{2,3}$-edge. The reference spectra
are taken from published papers: Fe$_2$O$_3$ \cite{kuepper04},
Fe$_3$O$_4$ \cite{kuepper04}, CoO \cite{de_groot93}, ZnCoO
\cite{barla:125201}, NiFe$_2$O$_4$ \cite{van_der_Laan_NiFeO} and
\CoFeO~\cite{laan:064407}.}\label{fig:XAS}
\end{figure}

The magnetic properties of 3$d$ transition metal elements, such as
Fe, Co and Ni are determined by the 3d valence electrons, which
can be investigated by L-edge XAS measurements (transition from
the 2$p$ shell to the 3$d$ shell). Fig. \ref{fig:XAS} shows the
L$_{2,3}$ XAS of Fe, Co and Ni in the two samples, measured in TEY
mode. The spectra of pure metals and some oxides are also shown
for comparison. The metal spectra mainly show two broad peaks,
reflecting the width of the empty $d$-bands, while the oxide
spectra exhibit a considerable fine structure of the $d$-bands,
the so-called multiplet structure. By comparison with
corresponding XAS of pure metals, one can qualitatively conclude
that metallic Fe, Co and Ni are not present in the samples. In
Fig. \ref{fig:XAS}(a) one can see the multiplet structure of Fe
L$_{2,3}$ XAS. The most noticeable feature is the rather
pronounced peak at the low energy part of the L$_3$ edge. This is
a common feature for ferrite materials \cite{Hochepied01}.
Multiplet calculations for FeO and $\alpha$-Fe$_2$O$_3$ reveal
that the shoulders at 705.5 eV and 718.5 eV [indicated by the
vertical arrows in the Fig. \ref{fig:XAS}(a)] are associated with
Fe$^{2+}$ ions \cite{PhysRevB.52.3143}. Note that these features
disappear in the spectrum of Fe$_2$O$_3$. Following these
arguments, the Fe ions in our samples are mainly Fe$^{3+}$ ions.

\begin{figure} \center
\includegraphics[scale=0.8]{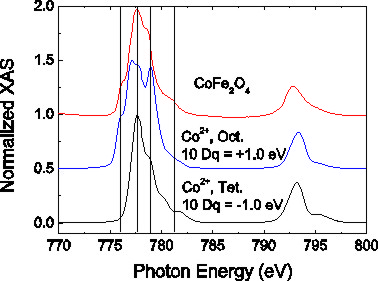}
\caption{The Co L$_{2,3}$-edge XAS spectrum of \CoFeO~along with
theoretical calculations for a tetrahedral (10 D$_q$ = -1 eV) and
an octahedral (10 D$_q$ = +1 eV) coordination of Co
ions..}\label{fig:Co_XAS_sim}
\end{figure}

The Co-L$_3$ edge [Fig. \ref{fig:XAS}(b)] is composed of a fine
structure with four features, a small peak at 775.5 eV, the total
maximum in absorption at 777 eV, followed by a shoulder at 778 eV
and a further satellite at 780 eV. Since in this sample the Co is
in pure Co$^{2+}$ configuration, we can compare the spectrum with
reference compounds namely CoO (spectrum taken from Ref.
\onlinecite{de_groot93}) and Zn$_{0.75}$Co$_{0.25}$O (spectrum
taken from Ref. \onlinecite{barla:125201}). Co$^{2+}$ ions are at
octahedral sites and at tetrahedral sites in CoO and
Zn$_{0.75}$Co$_{0.25}$O, respectively. From the comparison of the
overall shape and satellite structure our spectrum is more similar
to that of CoO, and also similar to the XAS of CoFe$_2$O$_4$
presented in Ref. \onlinecite{laan:064407}. We can conclude that
the major part of Co$^{2+}$ ions are at octahedral sites. In order
to confirm this conclusion, we performed simulations of the local
electronic structure around the Co$^{2+}$ ions by means of full
multiplet calculations using the TT-MULTIPLETS program
\cite{groot94,groot05}. The energy levels of the initial
($2p^63d^7$) and final absorption state ($2p^53d^8$) are
calculated by means of the corresponding Slater integrals which
are subsequently reduced to 80\% (corresponds to their atomic
values). Then a tetrahedral or octahedral crystal field was
considered using a crystal field parameter of 10 D$_q$ = -1 eV and
10 D$_q$ = +1 eV, respectively. Finally the calculated spectra
were broadened with the experimental resolution for comparison. As
displayed in Figure \ref{fig:Co_XAS_sim}, one can see that the
measured spectrum reasonably reproduces the features in the
simulated octahedral coordination.

Fig. \ref{fig:XAS}(c) shows the comparison of Ni L$_{2,3}$ with
that in \NiFeO~\cite{van_der_Laan_NiFeO}. In the paper of Van der
Laan \emph{et al.} \cite{van_der_Laan_NiFeO}, the spectrum can be
well simulated by considering Ni in an octahedral crystal-field
coordination, \emph{i.e.} Ni ions are fully at octahedral sites.
However, the difference between the two spectra is quite clear,
especially at the L$_2$-edge. It could be due to the fact that Ni
ions are partially located at tetrahedral sites.

\begin{figure} \center
\includegraphics[scale=0.8]{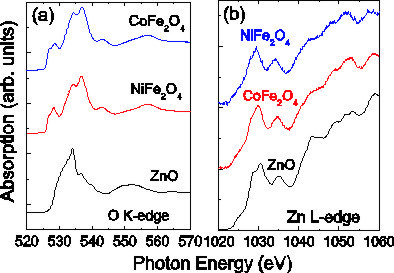}
\caption{The XA spectra of total electron yield (TEY) at the (a) O
K-edge, and (b) Zn L-edge.}\label{fig:XAS_OZn}
\end{figure}

The XAS spectra (Fig. \ref{fig:XAS_OZn}) were also recorded at the
O K-edge and Zn L-edge to prove the formation of ferrites and to
check if Zn ions are significantly incorporated into ferrites. All
shown spectra were measured in TEY mode, which is more sensitive
to the near surface-region where the ferrite nanocrystals were
formed. For the O K-edge, the difference between the \NiFeO,
\CoFeO~and ZnO is very clearly observed, which confirms the
coordination change of O ions. Actually the spectra in Fig.
\ref{fig:XAS_OZn}(a) are very similar to those of Fe$_{2}$O$_{3}$
and Fe$_{3}$O$_{4}$ \cite{gloter03}. Fig. \ref{fig:XAS_OZn}(b)
shows the comparison of the Zn L-edge spectra between ZnO embedded
with \NiFeO, \CoFeO~nanocrystals, and pure ZnO. The only
difference is that the fine structure in the spectrum of ZnO is
better resolved. This could be due to the lattice damage in ZnO by
ion implantation. No significant amount of Zn has been
incorporated into ferrites.

\subsubsection{XMCD}

Correspondingly, XAS recorded at 23 K in TEY (total electron
yield) mode at the Fe, Co and Ni absorption edge revealed a
pronounced dichroic behavior under magnetization reversal. XMCD is
a difference spectrum of two XA spectra, one taken with left
circularly polarized light, and the other with right circularly
polarized light.

The XMCD signal at the Fe L$_3$-edge XMCD for the two samples is
shown in Fig. \ref{fig:XMCD}(a) and (b). From the literature
\cite{takaobushi_PRB}, peak A is attributed to Fe$^{2+}$ at
octahedral sites, while peaks B and C are due to Fe$^{3+}$ at
tetrahedral and octahedral sites, respectively. By comparing the
relative height of peak A, B and C, we can draw some qualitative
conclusions on the cation site distribution in \NiFeO~and
\CoFeO~nanocrystals. Firstly, there are still some Fe$^{2+}$ ions
remaining, even if the ratio of implanted Ni(or Co) to Fe is
exactly 1:2. This could be due to the fact that Ni(or Co) and Fe
ions are not fully chemically reacted at the given annealing
condition. Relatively, there are more Fe$^{2+}$ ions at octahedral
sites in \CoFeO~than in \NiFeO. Secondly, part of Fe$^{3+}$ ions
are at tetrahedral sites in \NiFeO, while in \CoFeO~the Fe$^{3+}$
ions are mainly located at octahedral sites. Bulk \NiFeO~and
\CoFeO~are inverse spinels. The Ni$^{2+}$ and Co$^{2+}$ ions are
at octahedral sites, while half of the Fe$^{3+}$ ions are at
octahedral sites, and the other half are at tetrahedral sites.
With this ordering, the moments of Fe ions at octahedral and
tetrahedral sites cancel out, which results in a saturation
magnetization of 2$\mu_B$ per \NiFeO~formula unit
\cite{luders:134419}, and 3$\mu_B$ per \CoFeO~formula unit
\cite{antonov03}. However, in low dimensional spinels the cation
distribution is often different from bulk materials
\cite{chinnasamy00,luders:134419}. For the case of \NiFeO, if all
Ni$^{2+}$ replace the Fe$^{3+}$ at tetrahedral sites, resulting in
a normal spinel structure, the total magnetic moment can increase
up to 4$\mu_B$ per \NiFeO~formula. Therefore, the larger magnetic
moment in our \NiFeO~nanocrystals {as shown in Fig.
\ref{fig:MH_NiCoFeO} is probably due to a small amount of
Ni$^{2+}$ replacing the Fe$^{3+}$ at tetrahedral sites. This
cation distribution picture is in agreement with XAS analysis.
However, as discussed in section \ref{sec:mag}, one needs a
precise local concentration measurement to verify the ratio
$Fe:Ni=2:1$. Figs. \ref{fig:XMCD}(c) and (d) show the XMCD signal
at the L$_{2,3}$ edge for Ni and Co, respectively. They are
comparable with corresponding ferrites reported in literature
\cite{van_der_Laan_NiFeO,Hochepied01}. Note the fact that the
relative strength of the XMCD signal of \CoFeO~is much weaker than
that of \NiFeO. This is due to their different coercivity fields.
Due to the facility capability, a maximum field of 2000 Oe was
applied during XAS measurements. The saturation field of \CoFeO~is
much larger than that of \NiFeO~(see Fig. \ref{fig:MH_NiCoFeO}).

\begin{figure} \center
\includegraphics[scale=0.8]{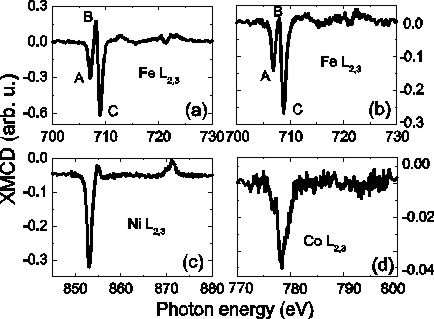}
\caption{XMCD at Fe, Ni and Co L$_{2,3}$ absorption edge. \NiFeO:
(a) and (c). \CoFeO: (b) and (d). Peak labels at Fe L$_3$-edge: A
for Fe$^{2+}$ at octahedral sites, B for Fe$^{3+}$ at tetrahedral
sites and C for Fe$^{3+}$ at octahedral sites
\cite{takaobushi_PRB}.}\label{fig:XMCD}
\end{figure}

\subsubsection{CEMS}

CEMS allows one to identify different site occupations, charge and
magnetic states of $^{57}$Fe. Fig. \ref{fig:CEMS} shows the CEM
spectra taken at room temperature for two samples, containing
\NiFeO~and \CoFeO~nanocrystals, respectively. The two samples
exhibit similar spectra. Using a least-squares computer program,
the spectra can be fitted well by three components. Two sets of
sextet hyperfine pattern and one doublet are resolved, all of
which are related to Fe$^{3+}$. The hyperfine parameters
calculated according to the evaluations of the spectra are given
in Table \ref{tab:cems_ferrite}. The outer sextet (S1) with a
larger magnetic hyperfine field corresponds to octahedral sites,
while the inner one (S2) with a smaller magnetic hyperfine field
to Fe$^{3+}$ at tetrahedral sites
\cite{ponpandian:192510,PhysRevB.54.9288,JJAP.40.4897,chakraverty:024115,subarna06}.
This feature of two sextets is a fingerprint that identifies
ferrites. The relative line intensities of the sextets differ from
those of a polycrystalline powder material indicating the presence
of a texture. Note that the magnetic hyperfine field is
considerably smaller than the values of around 50 T for typical
\NiFeO~or \CoFeO~\cite{JJAP.40.4897,subarna06}, which results from
the size effect \cite{chakraverty:024115}. The doublet (D) is a
more questionable component. Most probably it corresponds to
smaller ferrite nanocrystals, which are superparamagnetic at room
temperature. However, its isomer shift and electric quadrupole
splitting values are considerably larger than the values reported
in Ref. \onlinecite{subarna06}.

The cation distribution between the two sublattices generally
determines the magnetic properties of the spinel system and can be
calculated as a ratio between the relative areas of the respective
hyperfine field distributions. As shown in Table
\ref{tab:cems_ferrite}, there are more Fe$^{3+}$ ions at
octahedral sites for both samples. That means that the \NiFeO~and
\CoFeO~nanocrystals are not purely inverted ferrites and Ni or Co
ions partially occupy tetrahedral sites, which is in good agreement with the results of XAS. 

\begin{figure} \center
\includegraphics[scale=0.60]{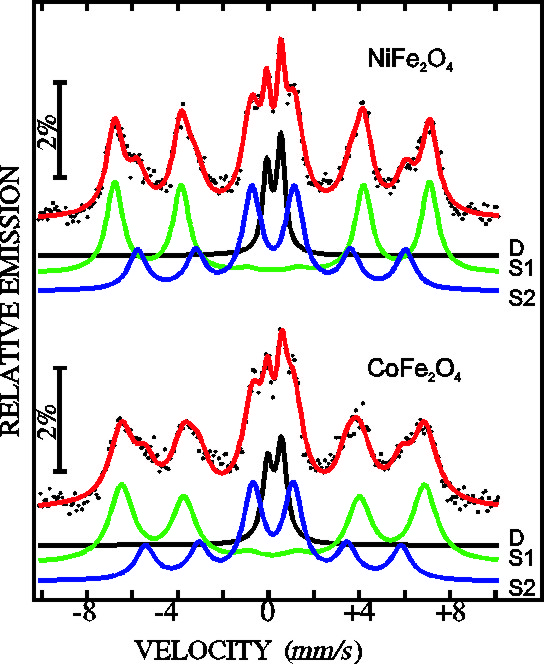}
\caption{Room temperature CEMS of \NiFeO/ZnO and \CoFeO/ZnO
composites. The notations for the fitting lines are given as D
(doublet) and S1, S2 (sextet).}\label{fig:CEMS}
\end{figure}

\begin{table*}
\caption{\label{tab:cems_ferrite}Hyperfine parameters measured
using CEMS for the two samples. The percentage of occupancies of
tetrahedral- and octahedral-sites by Fe$^{3+}$ ions. B$_{hf}$:
hyperfine field, A: relative area of each component, $\delta$:
isomer shift, $\Delta$: quadrupole spliting.}
\begin{ruledtabular}
\begin{tabular}{ccccc|cccc|ccc}
   & \multicolumn{4}{c} {S1 (octahedral)} & \multicolumn{4}{c} {S2 (tetrahedral)} & \multicolumn{3}{c} {D}\\
   & B$_{hf}$ & A & $\delta$ & $\Delta$ & B$_{hf}$ & A & $\delta$ & $\Delta$ & A & $\delta$ & $\Delta$ \\
Sample & (T) & (\%) & ($mm/s$) & ($mm/s$) & (T) & (\%) & ($mm/s$) & ($mm/s$) & (\%) & ($mm/s$) & ($mm/s$) \\
  \hline
  \NiFeO~& 43 & 47.7 & 0.26 & 0.03 & 36.5 & 39 & 0.28 & 0.09 & 13.3 & 0.34 & 0.63 \\
  \CoFeO~& 41.4 & 47.1 & 0.28 & 0.06 & 34.9 & 40.8 & 0.31 & 0.05 & 12.1 & 0.38 & 0.62 \\
\end{tabular}
\end{ruledtabular}
\end{table*}

\subsection{Magneto-transport properties}

Note that both bulk \NiFeO~and \CoFeO~are insulators with resistivity of 10$^2$- 10$^3$ $\Omega$cm at room
temperature \cite{luders_AM,snoeck:104434}. The resistivity of \NiFeO~single crystals monotonically increases
with decreasing temperature \cite{zalesski00}. However, the corresponding thin film materials can be rather
conductive \cite{luders_AM}. In ref. \onlinecite{luders_AM}, the authors show that the \NiFeO~films grown in
pure Ar atmosphere have a room temperature resistivity three orders-of-magnitude smaller ($\rho$ around 100
m$\Omega$cm). The temperature dependence $\rho$(T) is similar to that of magnetite. On the other hand, ZnO
single crystals grown by the hydrothermal method show a high bulk and surface resistivity, with the bulk
conduction dominated by a deep donor \cite{kassier:014903}. Typically, the free charge carrier concentration
amounts to $1\times 10^{14}$ cm$^{-3}$ and the mobility to 200 cm$^2$V$^{-1}$s$^{-1}$ (Ref. \onlinecite{maeda}).
We measured the temperature dependence of the sheet resistance of the composites of \NiFeO~and
\CoFeO~nanocrystals and ZnO from 20 or 40 to 290 K. Fig. \ref{fig:RT}(a) shows the Arrhenius plot, the sheet
resistance $R_{s}$ on a logarithmic scale as a function of reciprocal temperature. Note that the resistivity of
both samples is below 0.1 $\Omega$cm at room temperature assuming a thickness of 80 nm, which is three orders of
magnitude smaller than that of bulk ferrites or ZnO \cite{maeda}. The resistance/resistivity of composites of
\CoFeO~and ZnO is one order smaller than that of \NiFeO~and ZnO. The amount of n-type defects in ZnO created by
means of implantation and annealing is expected to be similar. Therefore, the larger conductivity in the
composite of \CoFeO~and ZnO is due to the mixing of Fe$^{2+}$ and Fe$^{3+}$ ions at octahedral sites
\cite{luders_AM}. The temperature dependence of the resistance is more or less the same for both samples. Two
different regimes are found. One is the high temperature part (above 150 K), where the resistance slightly
decreases with decreasing temperature. This is a hint of metallic character. However, the electron concentration
(around $6\times10^{18}$ cm$^{-3}$ assuming a thickness of 80 nm) as shown in Fig. \ref{fig:RT}(b) is far below
the critical value ($4\times10^{19}$ cm$^{-3}$) of the metal-insulator transition in n-ZnO \cite{xu:205342}. In
Refs. \onlinecite{luders:134419,luders_AM}, a metallic electrical conductivity has been obtained in ultrathin
\NiFeO~films, which is attributed to an anomalous distribution of the Fe and Ni cations among tetrahedral and
octahedral sites. Therefore, we attribute the metallic character in our samples to the presence of \NiFeO~(or
\CoFeO) nanocrystals. The second regime is in the temperature range below 150 K. In this regime, the samples
show a semiconducting conductivity. The thermal activation energy $E_a$ of free carriers can be determined
according to the following equation:
\begin{equation}\label{RT}
    \rho=e^{\frac{E_a}{k_BT}}+R_{s0},
\end{equation}
where $k_B$ is the Boltzmann constant and $R_{s0}$ a temperature
independent contribution to the resistivity. In Fig.
\ref{fig:RT}(a) the solid lines show the fitting, resulting in a
thermal activation energy of $\sim$28 meV for both samples. A
similar thermal activation energy of 21 meV has been found in
hydrothermally grown ZnO single crystals after high-temperature
annealing \cite{kassier:014903}. At low temperatures the
impurities freeze out. In Ref. \onlinecite{zalesski00}, the
authors show that in their measured temperature range from around
77 K to room temperature the \NiFeO~single crystals exhibit
semiconducting conductivity, and the activation energy is around
60 meV. Fig. \ref{fig:RT}(b) displays the temperature dependent
carrier concentration and Hall mobility. The sheet electron
concentration increases with temperature and reaches to
4.8$\times10^{13}$ cm$^{-2}$. Its temperature dependence can be
well fitted by the function $e^{-E_a/k_BT}$. Fig. \ref{fig:RT}(b)
also shows the temperature-dependent mobility. The electron
mobility $\mu$ reaches a maximum of above 900 $cm^2/Vs$ at 65 K.
Actually such large electron mobility and concentration were also
observed in ion implanted ZnO \cite{potzger:023510} and in virgin
ZnO annealed in N$_2$ \cite{look:122108}.

\begin{figure} \center
\includegraphics[scale=0.62]{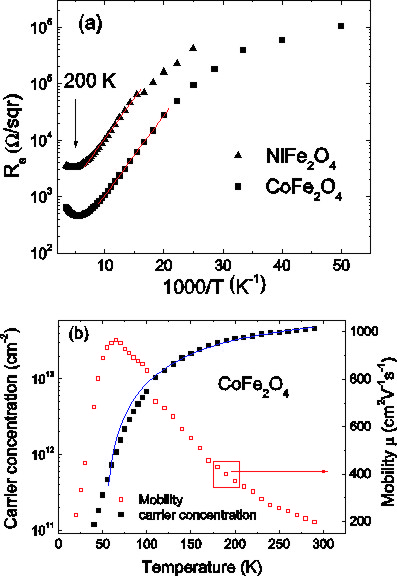}
\caption{(a) The temperature dependent sheet resistance (the solid
lines show the linear fitting in the temperature range of 60-100
K), and (b) free carrier concentration and mobility of the
composites of \NiFeO~or \CoFeO~nanocrystals and ZnO. The solid
line is a fitting of carrier concentration by the function
$e^{-E_a/k_BT}$.}\label{fig:RT}
\end{figure}

We also measured the magnetic field dependent resistance (MR) for
the composites of \NiFeO~(\CoFeO) nanocrystal and ZnO as shown in
Fig. \ref{fig:MR_NiCoFeO}. MR is defined as
\begin{equation}\label{MR}
    MR=(R[H]-R[0])/R[0].
\end{equation}
The two samples exhibit a similar MR behavior. Only positive MR
has been observed, and MR decreases quickly from around 16\%(6 T)
at 20 or 40 K to 0.2\% (6 T) at 290 K. The overall shape of the
field dependent MR is quadratic, and shows no sign of saturation.
We attribute it to ordinary MR effect resulting from the curving
of the electron trajectory due to Lorenz force in a magnetic
field. The characteristic quantity is the Landau orbit,
$L_H=(eH/{\hbar}c)^{-1/2}$, which is temperature independent.
Another parameter is the dephasing length $L_{Th}$ of electrons,
the diffusing distance between two elastic scattering events,
which decreases with increasing temperature. When the dephasing
length is much smaller than the Landau orbit,
${L_{Th}}^2/{L_H}^2{\ll}1$, the magnetoresistance is quadratic and
non saturating. Actually the field dependent MR can be fitted well
as a H$^2$ dependence (not shown). That means the dephasing length
is very small in this sample due to the presence of \NiFeO~or
\CoFeO~nanocrystals. In the literature a large positive MR up to
several hundreds or thousands percent has been observed in
regularly ordered nanowires \cite{liu:1436} or nanocolumns
\cite{jamet06}. A non-saturating positive MR effect is expected to
be useful for wide-range field sensing. A positive MR has also
been observed in Co-doped ZnO films, and modelled by considering
$s-d$ exchange \cite{xu:205342,xu:134417}. Note that the MR at 20
K in Fig. \ref{fig:MR_NiCoFeO} exhibits a small contribution
indicated by the arrows, which saturates at low fields. This
contribution could be due to $s-d$ exchange considering that a
small amount of Co$^{2+}$ or Ni$^{2+}$ ions remains in a diluted
state. On the other hand, no negative MR has been observed, which
often was found in the hybrid system of MnAs and GaAs
\cite{wellmann98,akinaga98}.

\begin{figure} \center
\includegraphics[scale=0.62]{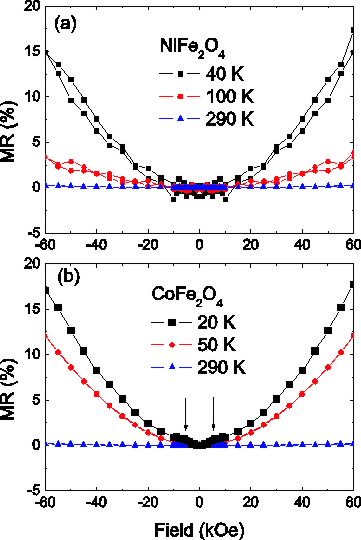}
\caption{The field dependent MR of (a) the composites of
\NiFeO~nanocrystals and ZnO and (b) the composites of
\CoFeO~nanocrystals and ZnO. The solid lines are guides to the
eye.}\label{fig:MR_NiCoFeO}
\end{figure}

The Hall resistivity
\begin{equation}\label{Hall}
\rho_{xy}=R_HB+R_M\mu_0M
\end{equation} is known to be a sum of the ordinary and anomalous Hall terms, where $B$ is
magnetic induction, $\mu_0$ magnetic permeability, $M$
magnetization, $R_H$ the ordinary Hall coefficient, and $R_M$ the
anomalous Hall coefficient. The ordinary and anomalous Hall term
is linear in B and M, respectively. After subtracting the linear
part, the ordinary Hall effect, a clear AHE also has been observed
in the two samples, as shown in Fig. \ref{fig:AHE_NiCoFeO}. AHE
vanishes at temperatures above 100 K. Obviously the AHE curve does
not coincide with the magnetization curve as shown in Fig.
\ref{fig:MH_NiCoFeO} and Fig. \ref{fig:T_MT_NiCoFeO}. It is
difficult to correlate the observed AHE to \NiFeO~or
\CoFeO~nanocrystals. Usually, AHE is not expected or very weak for
a semiconductor with embedded magnetic nanoparticles
\cite{shinde04,zhang:085323}. If one considers that the shape of
the AHE curves mimics the M-H curves, the AHE is likely due to
some paramagnetic contributions or magnetism induced by intrinsic
defects \cite{xu:082508}.

\begin{figure} \center
\includegraphics[scale=0.62]{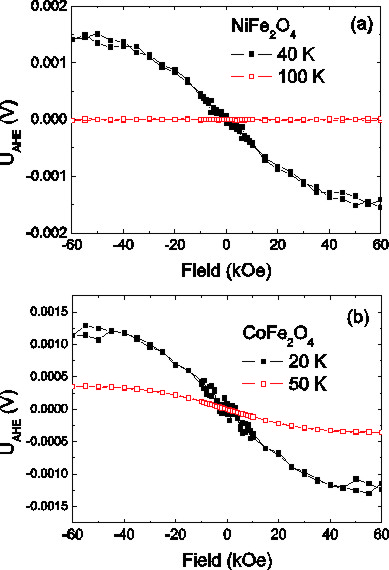}
\caption{Anomalous Hall voltage vs magnetic field for (a) the
composites of \NiFeO~nanocrystals and ZnO and (b) the composites
of \CoFeO~nanocrystals and ZnO.}\label{fig:AHE_NiCoFeO}
\end{figure}

\section{Conclusions}

(I) Nano-scaled ferrite materials attract considerable research
attention due to their cation distribution and applications as
dielectric materials
\cite{chinnasamy00,PhysRevB.60.3400,bohra:262506,thakur:262501}.
Usually, ferrite nanoparticles are formed by mechanical or
chemical methods. We have demonstrated the formation of \NiFeO~and
\CoFeO~nanocrystals inside a ZnO matrix. Ion beam synthesis has
its own obvious advantage of allowing lateral patterning
\cite{Fassbender3M}.

(II) \NiFeO~and \CoFeO~nanoparticles are crystallographically
oriented with respect to the ZnO matrix. They show similar
structural properties, but different magnetic, and transport
properties. Considering the rich phases of spinel ferrites
(TMFe$_2$O$_4$, TM=Ni, Co, Fe, Mn, Cu, Zn), our results
demonstrate the possibility to have a new magnet/semiconductor
hybrid system. This system can be tuned over a large variety of
magnetic and transport properties. However, the observed MR and
AHE are likely not related to the presence of \NiFeO~and
\CoFeO~nanocrystals. This could be due to the imperfect interface
between nanocrystals and ZnO matrix. This problem could be solved
by epitaxial growth methods, \eg, pulsed laser deposition. A
multilayered structure of ferrites/ZnO could be grown, and opens a
path towards semiconducting spintronic devices.

(III) Our results suggest the possible integration of ferrites
with semiconducting ZnO, which would allow the integration of
microwave with semiconductor devices. The combination of ferrites
and conventional semiconductors, \eg, Si and GaAs, proves to be
challenging due to the requirements of oxygen atmosphere and high
temperature for ferrites \cite{chen_APL_ferrites}.

\section{Acknowledgement}

The authors (S.Z., Q.X. and H.S.) thank financial funding from the
Bundesministerium f\"{u}r Bildung und Forschung (FKZ03N8708). Q.
X. is supported by the National Natural Science Foundation of
China (50802041). The Advanced Light Source is supported by the
Director, Office of Science, Office of Basic Energy Sciences, of
the U.S. Department of Energy under Contract No.
DE-AC02-05CH11231.

During the press of this manuscript, we observed that an all-oxide ferromagnet/semiconductor (Fe$_3$O$_4$/ZnO)
heterostructure has been realized by other groups using pulsed laser deposition \cite{nielsen:162510}.


\end{document}